\documentclass[manuscript]{aastex63}  
\usepackage{txfonts}
\usepackage{latexsym,bm}
\usepackage{url}
\usepackage{color}
\usepackage{colortbl}
\usepackage{multirow}
\usepackage{ulem} 

\definecolor{mygray}{gray}{.9}

\shorttitle{electron acceleration at the shock}

\shortauthors{Kong and Qin}

\begin{document}

\title{Suprathermal electron acceleration by a quasi-perpendicular shock: 
simulations and observations} 

\correspondingauthor{G. Qin}
\email{qingang@hit.edu.cn}

\author{F.-J. Kong}
\author[0000-0002-3437-3716]{G. Qin}

\affiliation{School of Science, Harbin Institute of Technology, Shenzhen,
518055, China; qingang@hit.edu.cn}

\begin{abstract}
The acceleration of suprathermal electrons in the solar wind is 
mainly associated with shocks driven by interplanetary
coronal mass ejections (ICMEs). 
It is well known that the acceleration of electrons is much more efficient at 
quasi-perpendicular shocks than at quasi-parallel ones. Yang et al. (2018, ApJ,
853, 89) (hereafter YEA2018) studied the acceleration of suprathermal
electrons at a quasi-perpendicular ICME-driven shock event to claim the
important role of shock drift acceleration (SDA). Here, we perform 
test-particle simulations to study the acceleration of electrons in 
this event, by calculating the downstream electron intensity 
distribution for all energy
channels assuming an initial distribution based on the averaged upstream 
intensities. We obtain simulation results similar to the observations from
YEA2018 as follows. It is shown that the ratio of downstream
to upstream intensities peaks at about 90$^\circ$ pitch angle. In addition,
in each pitch angle direction the downstream electron energy spectral index
is much larger than the theoretical index of diffusive shock acceleration.
Furthermore, considering SDA, the estimated drift length is proportional to
the electron energy but the drift time is almost energy independent. 
Finally, we use a theoretical model based on SDA to describe the drift
length and time, especially, to explain their energy dependence. These
results indicate the importance of SDA in the acceleration of 
electrons by quasi-perpendicular shocks.
\end{abstract}


\section{INTRODUCTION}

Shock acceleration is a crucial source for energetic particles in the 
heliosphere and galaxcy. 
Charged particles gain energy via different mechanisms at shocks. The first one
is first-order Fermi acceleration (FFA) due to the relative motion
of scattering centers in the upstream and downstream regions. The second one is 
shock drift acceleration (SDA) as a result of gradient drift along the direction of 
convective electric field. The third one is stochastic acceleration 
(second-order Fermi) associated with the downstream turbulence.
The well-known diffusive shock acceleration (DSA) theory
\citep{Axford1977,Krymsky1977,Bell1978,Blandford1978}, which successfully 
explains the power-law distribution of accelerated particles observed
in the universe,  is the combination of FFA and SDA.

Various heliospheric shocks, such as planetary bow shocks, coronal shocks,
propagating interplanetary shocks, and the solar wind termination shock, are
excellent acceleration sites of particles. It is assumed that there are two kinds
of solar energetic particle (SEP) events observed at 1 au, one is impulsive
events originating in solar flares 
\citep[e.g.,][]{Cliver1982,Mason1984,Cane1986}, the other is gradual events
associated with interplanetary shocks driven by coronal mass ejections (CMEs) 
\citep[e.g.,][]{Kahler1978,Kahler1984,Reames1999}. Most space weather disturbances
can be traced to large and long-duration gradual SEP events.
Particle acceleration by CME-driven shocks has been widely studied
\cite[e.g.,][]{Reames1996,Zank2000,Li2003,Rice2003,Desai2008, Kong2017, Kong2019b, 
Qin2018}; however, 
some problems still remain unresolved concerning particle acceleration at
propagating interplanetary shocks. The geometry on the shock surface is variable
when a CME-driven shock propagates from the sun into the interplanetary space.
In addition, the turbulence level and
other solar wind conditions are changeable. Therefore it is important to study the
shock acceleration efficiency with varying obliquity angle, turbulence level, 
and other conditions \citep[e.g.,][]{Giacalone2005,Guo2015,Qin2018}. 
It is suggested by \citet{Qin2018} that particle acceleration processes
with weak scatterings (weak turbulence) are generally FFA for parallel 
(quasi-parallel) shocks and SDA for perpendicular (quasi-perpendicular) shocks,
and that SDA is more efficient than FFA.

Recently, some interesting phenomena are observed in the acceleration of electrons 
by shocks. On the one hand, energetic electrons can be produced by shock
acceleration near the sun.
\citet{Holman1983} suggested that electrons are accelerated to high energies
in the solar corona by SDA. Energetic electrons, which are observed as solar type 
$\text{\uppercase\expandafter{\romannumeral2}}$ radio bursts, can also be generated
via resonant interaction with whistler waves 
at quasi-perpendicular shocks in the solar corona \citep{Miteva2007}. 
 Furthermore, \citet{Li2013} and \citet{Kong2013}
studied electron spectral hardening in
solar flares with observations from Gamma-Ray Spectrometer (SMM) onboard \emph{Solar 
Maximum Mission (SMM)}, to suggest that energetic electrons are produced
by SDA. 

On the other hand, there exists 
low-energy electrons in interplanetary (IP) space such as suprathermal electrons, which, 
to constitute important components of solar wind halo, strahl, and superhalo populations 
\citep[e.g.,][]{Feldman1975,Lin1998,Maksimovic2005},
can be accelerated by following the meandering 
magnetic field lines repeatedly across the shock \citep{Jokipii2007, Guo2010}.
\citet{Kajdic2014} 
found 90$^\circ$ pitch angle enhancements of suprathermal electrons at IP shocks measured
by Solar Wind Electron Analyzer (SWEA) onboard \emph{STEREO-A}. 
 \citet[hereafter YEA2018]{Yang2018} investigated 
two strong electron flux enhancement events measured by electron electrostatic
analyzers (EESA) in the 3DP instrument onboard \emph{Wind}, one with a
quasi-perpendicular shock, the other one with a quasi-parallel shock.
It is found that at energies of $\sim$0.4--50 keV, the ratios of the downstream to
ambient electron intensities all peak at around 90$^\circ$ pitch angle. Besides, the
energy spectrum in each pitch angle direction in the downstream can be fit to a double
power-law with a spectral index much larger than the theoretical one from DSA. They
thus assumed that SDA is the dominant acceleration mechanism in both the shock
events, and they obtained that the drift length
is roughly proportional to the electron energy but the drift time almost does not vary
with energy.

In \citet{Kong2019a} we performed numerical simulations to obtain 90$^\circ$ pitch angle
enhancements for three sample energy channels in the range of 89--257 eV at a 
quasi-perpendicular shock on 24 April 2008 that was studied by 
\citet{Kajdic2014}.
In this paper, we study numerically the acceleration of suprathermal electrons 
in the range of $\sim$0.3--40 keV at the quasi-perpendicular shock on 2000 
February 11 that was studied by YEA2018. Pitch angle distributions (PADs)
for all of the 12 energy 
channels, energy spectra for the parallel, perpendicular, and anti-parallel 
directions, and spectral indices for all pitch angles are obtained by solving
the motion equation of electrons using a backward-in-time test-particle method.
The electron drift length and drift time are also
estimated using the method of unchanged phase space density after the acceleration
of electrons (YEA2018). Finally, a theoretical model is obtained to describe the
drift length and drift time. In Section 2 we briefly introduce the instruments onboard
\emph{Wind} used in the study, and we list the shock parameters for the event.   
We describe our physical and numerical models to accelerate electrons
in Section 3. In Section 4, we first show the estimation of electron
drift length and drift time from the data of distribution functions following YEA2008, then we
provide a theoretical model of drift length and drift time. 
In Section 5 the simulation results and comparisons with 
observations are presented. Finally, we show in Section 6 the discussion 
and conclusions.

\section{Observations}

The $Wind$ spacecraft was launched on November 1, 1994, and located around
the L1 Lagrangian point. The 3DP instrument onboard \emph{Wind} is
designed to measure the distribution of suprathermal electrons and ions
in the solar wind. The electron electrostatic analyzers (EESA) in the 
3DP instrument measures the energy ranges of $\sim$3--1000 eV and 
$\sim$0.1--30 keV, respectively, for the low and high energy detectors.
The 3DP instrument provides three-dimensional data with eight pitch
angle channels of $22.5^\circ$ interval.
For more details of \emph{Wind} 3DP instrument see \citet{Lin1995}.
The magnetic field data are measured by the magnetic field instrument (MFI)
\citep[see,][]{Farrell1995}, and the plasma parameters in the solar wind such as 
bulk flow speed $V_{\text{sw}}$ and proton number density $n_{\text p}$ are 
provided by the Solar Wind Experiment (SWE) instrument \citep{Ogilvie1995}. 

As already mentioned above, we focus on the acceleration of electrons by 
the quasi-perpendicular shock on 2000 February 11. 
The shock arrived at \emph{Wind} at 23:34 UT according to \emph{Wind} IP
shock list \url{https://www.cfa.harvard.edu/shocks/wi_data/}, 
with a shock-normal angle of $\theta_{\text{Bn}}\sim89^\circ$, a shock speed
of $V_{\text{sh}}\sim682~\text{km s}^{-1}$, and a compression ratio 
$s\sim2.87$ from calculations of YEA2018. The upstream magnetic 
field, solar wind bulk speed, and proton number density are set to be 
$B_{01}=7.0~\text{nT}$, 
$V_{\text{sw}}=434~\text{km s}^{-1}$, and $n_{\text p}=5.19~\text{cm}^{-3}$,
respectively, 
by averaging the observational data over the time range of 23:20--23:30 
(data used from the website \url{https://cdaweb.sci.gsfc.nasa.gov}).
The upstream speed is set to be $U_1=V_{\text{sh}}-V_{\text{sw}}\sim
248~\text{km s}^{-1}$ for simplicity. We obtain the upstream Alfv$\acute{\text e}$n
speed $V_{\text A1}=67~\text{km s}^{-1}$, and Alfv$\acute{\text e}$n 
Mach number $M_{\text A1}=3.70$.

\section{IP shock models to accelerate electrons}

We study here the acceleration of electrons at an IP shock by solving numerically 
the equation of motion of test particles, and such method has been used in previous studies
\citep{Decker1986a,Decker1986b,Giacalone2005,Giacalone2009,Kong2017, Kong2019b, 
Qin2018}.

\subsection{Physical Model}

For simplicity, we consider a planar shock with the geometry shown in the cartoon 
of Figure 1 in \citet{Kong2017}. The shock is located at $z=0$ with a thickness 
$L_{\text{th}}$. The plasma flows in the positive $z$-direction
with the upstream and downstream speeds, $U_1$, $U_2$, respectively, in the shock frame.
In this work, we usually use subscripts $1$ and $2$ to indicate the upstream and
downstream, respectively.
In the shock transition the plasma speed is assumed to be in the form of
\begin{equation}
U(z)=\frac{U_1}{2s}\left\{\left(s+1\right)+\left(s-1\right)\tanh\left[\tan\left(-\frac{\pi
z}{L_{\text{th}}}\right)\right]\right\},
\end{equation}
where $s$ is the shock compression ratio. The motion equation of test particles is given by
\begin{equation}
\frac{d\boldsymbol p}{dt}=q[\boldsymbol E(\boldsymbol r,t)+\boldsymbol v\times\boldsymbol B(\boldsymbol r,t)],    
\end{equation}
where $\boldsymbol p$ is the particle momentum, $\boldsymbol v$ is the particle velocity, $q$ is the electron charge, 
t is time. The electric field $\boldsymbol E$ is the convective electric field 
$\boldsymbol E=-\boldsymbol U\times\boldsymbol B$. The total magnetic field consists of 
the background magnetic field and turbulent magnetic field, and is given by
\begin{equation}
\boldsymbol B(x^\prime,y^\prime,z^\prime)=\boldsymbol B_0+\boldsymbol b(x^\prime,y^\prime,z^\prime).
\end{equation} 
Note that the background magnetic field $\boldsymbol B_0$ is in the $x-z$ plane. 
The input parameters for the shock are shown in Table \ref{shockpara}. 
The value of shock thickness $L_{\text{th}}$ is set as $2\times10^{-6}$ au 
according to YEA2018.

The turbulent magnetic field is given by
\begin{equation}
\boldsymbol b(x^\prime,y^\prime,z^\prime)=\boldsymbol b_{\text{slab}}(z^\prime)+\boldsymbol b_{\text{2D}}(x^\prime,y^\prime),
\end{equation}
where $\boldsymbol b$ is a turbulent magnetic field perpendicular to 
$\boldsymbol B_0$ with zero mean, and $(x^\prime, y^\prime, z^\prime)$ is 
the coordinate system with $z^\prime$ in the direction of $\boldsymbol B_0$.  
The turbulent magnetic field $\boldsymbol b$ is composed of slab and two-dimensional (2D) 
components with the energy density ratio assumed to be
$E_{\text{slab}}:E_{\text{2D}}=20:80$
\citep[``two-component" model,][]{Matthaeus1990,Zank1992,Bieber1996,Gray1996,Zank2006}.
We assume the slab correlation length $\lambda=0.02~\text{au}$ at 1 au, and the 2D 
correlation length $\lambda_x=\lambda/2.6$ on the basis of previous studies
\citep{Osman2007,Weygand2009,Weygand2011,Dosch2013}. A dissipation range in which 
low-energy electrons resonate is included in the slab turbulence, as applied in 
the work of \citet{Qin2018}. The break wavenumber $k_{\text b}$ from the inerial range to
the dissipation range is assumed to $k_{\text b}=10^{-6}~\text{m}^{-1}$ based on the observational
investigations in \citet{Leamon1999}. The values of spectral indices of the 
inertial and dissipation ranges ($\beta_{\text i}=5/3,~\beta_{\text d}=2.7$) are set as the same as
those in \citet{Qin2018}. A periodic turbulence box with sizes $10\lambda\times10\lambda$ 
and $25\lambda$ for the 2D and slab components, respectively, is adopted in the simulations.
The turbulence levels of the upstream and downstream regions are taken to be 
$(b/B_0)^2=0.25$ and $0.36$, respectively. The input parameters for the turbulence
are shown in Table \ref{turbupara}.

\subsection{Numerical Model}

Based on the \emph{Wind}/3DP observations, we simulate pitch angle distributions (PADs) 
of 12 energy channels with central energies $\sim 0.266,~ 0.428,~ 0.691,~ 1.116,
~1.952, ~2.849, ~4.161, ~6.076, ~8.875, ~12.96, ~27.32$, and $39.50$ keV. A 
backward-in-time test-particle method is used to simulate the PAD of a given energy 
channel $E_i$ ($i=1,2,3,...,12$) in the downstream of the shock. A total number of 
30, 000 electrons with an energy $E_i$ and a pitch angle $\mu_j$ are put 
into the downstream range $[z_0,z_1]$ at the initial time $t=0$, where
$z_0=L_{\text{th}}$ and $z_1=V_{\text{sh}}\Delta t\approx 2.7\times 10^{-3}$ au with
$\Delta t=10$ min. We take the spatial domain size
in the $x$, $y$, $z$ directions to $x_{\text{box}}=y_{\text{box}}=10^4\lambda$ and
$z_{\text{box}}=10^3\lambda$. The trajectory of each electron is followed using an 
adaptive 
step fourth-order Runge-Kutta method with a normalized accuracy to $10^{-9}$ until
the simulation time $t_{\text{acc}}=10$ min. After the numerical calculations, a few
electrons whose energy is less than $0.1E_i$ are discarded. The downstream pitch angle 
distribution, $f_{\text{dn}}(E_i,\mu_j)$, for $E_i$ channel can be obtained as
\begin{equation}
f_{\text{dn}}(E_i,\mu_j)=\frac{1}{N_{ij}}\sum^{N_{ij}}_{k=1}f_0(E_{ik},\mu_{jk}),
\end{equation}
where $N_{ij}$ is the number of test particles in the statistics, 
$f_0(E_{ik},\mu_{jk})$
is the initial distribution, and $E_{ik}$ and $\mu_{jk}$ are the $k$th particle 
energy and pitch angle, respectively, when it is traced back to the initial
time. The initial distribution is constructed 
by averaging the 3DP data in the time period of 23:20--23:30 before the shock arrival.
Note that we employ linear interpolation in log-log space between the adjacent 
particle energies to calculate the value of $f_0(E_{ik},\mu_{jk})$.
In addition, in this work, to use test-particle method, 
we do not consider wave excitation by the accelerated particles. 

\section{Drift length and time}
\subsection{Estimation of electron drift length and time from distribution functions}
In this subsection, following YEA2008, we show the estimation of the electron
drift length and drift time. To consider SDA, according to Liouville’s theorem, it is 
assumed that electrons remain the same phase space density after they
are accelerated from the upstream to downstream, i.e., 
\begin{equation}
    f_2(p_2)=f_1(p_1),
    \label{eq:same_phase_space_density}
\end{equation}
where $p$ and $f$ 
are the electron momentum and phase space density, respectively. Here, the
subscripts $1$ and $2$ mean the upstream and downstream of the shock, 
respectively.  The energy gain $\Delta E$ after the acceleration of upstream 
electrons with a momentum $p_1$ can be obtained considering the same phase space 
density. The electron drift length is then written as 
\begin{equation}
    L_{\text{drift}}=\frac{\Delta E}{q|\boldsymbol{E}|},
    \label{eq:Ldrift}
\end{equation} 
where $\boldsymbol{E}$ is the 
convection electric field. 

According to \citet{Jokipii1982}, the gradient drift 
velocity at the shock front is 
\begin{equation}
\boldsymbol{V}_{\text{drift}}=\boldsymbol{\hat{e}}_y 
\frac{pv}{3q}\left(\frac{B_{x1}}{B_1^2}
-\frac{B_{x2}}{B_2^2}\right)\delta(z),
\end{equation}
where $B_1$ and $B_2$ are the background magnetic fields with their $x$-components
$B_{x1}$ and $B_{x2}$ in the upstream and downstream, respectively. 
We can integrate the above equation for $z$ from $-L_{\text{th}}/2$ to
$L_{\text{th}}/2$ to obtain the average gradient drift velocity
\begin{equation}
\boldsymbol{\bar{V}}_{\text{drift}}=\boldsymbol{\hat{e}}_y 
\frac{pv}{3qL_{\text{th}}}\left(\frac{B_{x1}}{B_1^2}
-\frac{B_{x2}}{B_2^2}\right).
\label{eq:vdrift}
\end{equation}
In addition, the drift time, $T_{\text{drift}}$, can be obtained by 
\begin{equation}
T_{\text{drift}}=\frac{L_{\text{drift}}}{\bar{V}_{\text{drift}}}.
	\label{eq:Tdrift}
\end{equation}

\subsection{A theoretical model for electron drift length and time}
Next, we provide a theoretical model based on shock drift acceleration to
describe the drift length and drift time. Since the electron 
gyroradius is much smaller than the shock thickness $L_{\text{th}}$, it is
assumed that
an electron can be accelerated by SDA when it is in the quasi-perpendicular
shock transition range. 
On the other hand, when particles are in the shock transition range they would
move the downstream as the fluid convection, so the electron drift time 
$T_{\text{drift}}$ can be written as
\begin{equation}
    T_{\text{drift}}= \frac{L_{\text{th}}}{2U_1}+
	\frac{L_{\text{th}}}{2U_2},
        \label{eq:Tdrift_th}
\end{equation}
where $U_i$ indicates the fluid speed upstream and downstream of the shock with $i=1$
and $i=2$, respectively. Since the shock thickness $L_{\text{th}}$ and fluid
convection speed $U_i$
can be considered constant, the drift time $T_{\text{drift}}$ is constant too,
\begin{equation}
    T_{\text{drift}}\propto E^{0}.
    \label{eq:Tconst}
\end{equation}

Furthermore, according to Equations (\ref{eq:vdrift}), (\ref{eq:Tdrift}), and  
(\ref{eq:Tdrift_th}), the electron drift length $L_{\text{drift}}$ can be calculated as
 \begin{equation}
     L_{\text{drift}}=T_{\text{drift}}\bar{V}_{\text{drift}}=
\frac{pv}{6q}\left(\frac{1}{U_1}+
	\frac{1}{U_2}\right)
\left(\frac{B_{x1}}{B_1^2}
-\frac{B_{x2}}{B_2^2}\right).
    \label{eq:Ldrift_th}
 \end{equation}
It is found that the electron drift length $L_{\text{drift}}$ is proportional
to $E$ if the relativistic effects are not considered, i.e.,
 \begin{equation}
     L_{\text{drift}}\propto E.
 \end{equation}

\section{Simulation results and the comparisons with observations}

\begin{figure}
\epsscale{1.}
\plotone{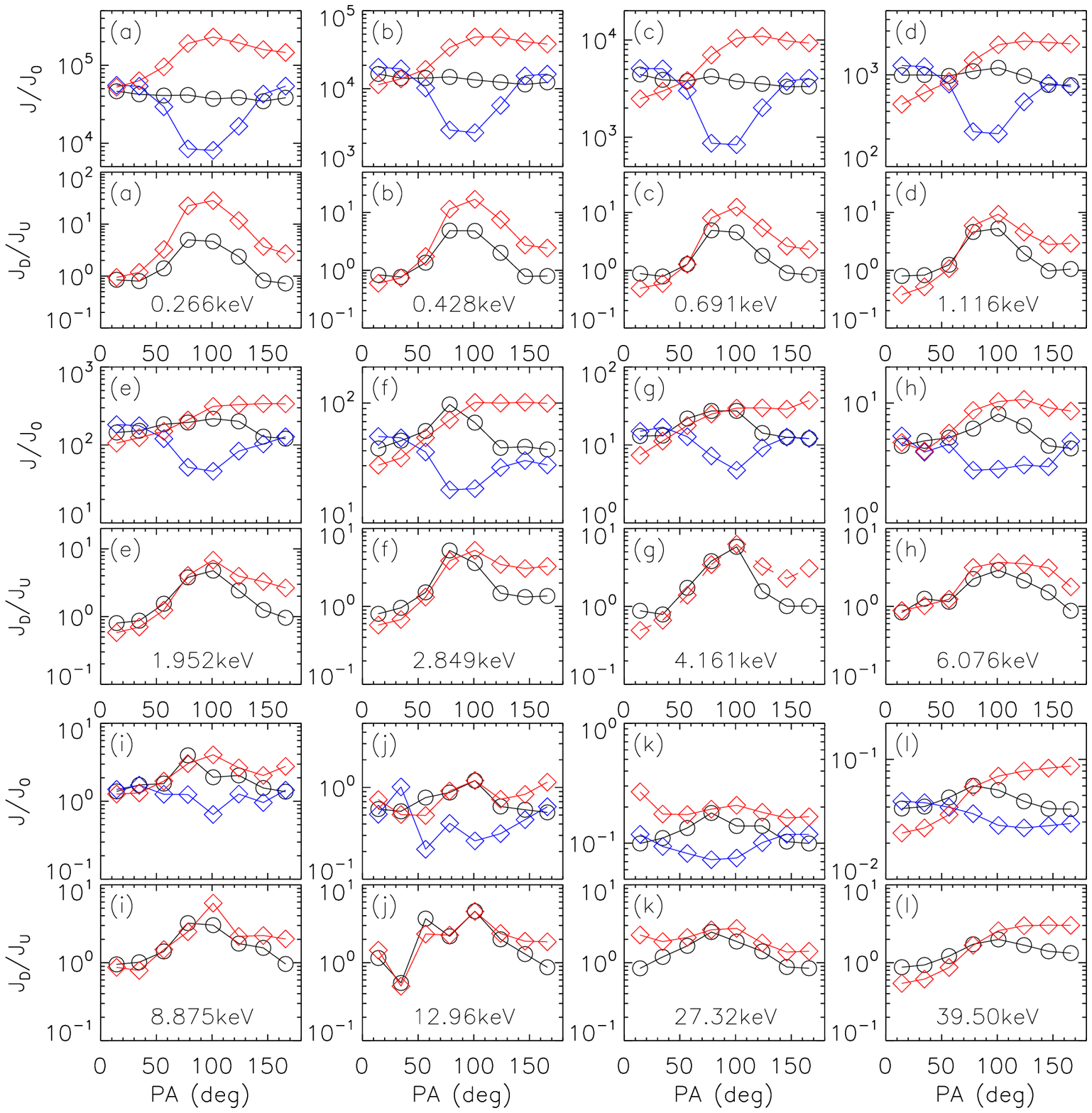}
        \figcaption{Electron differential intensity vs. pitch angle in the energy 
channels of 0.266--39.50 keV marked with (a)--(l). In the upper panels of the 
figures (a)--(l), the blue 
and red diamonds correspond to 10-minute averaged upstream and downstream intensities
from observations, respectively, and the black circles are results from simulations 
using an initial distribution based on the observed upstream intensities. 
In the lower panels, red diamonds 
(black circles) show the ratio of the downstream intensities for 
observations (simulations) to the upstream intensities.
Note that $J_0$ denotes a unit of intensity (1 cm$^{-2}$s$^{-1}$sr$^{-1}$eV$^{-1}$).    
\label{fig:fluxPA}}
\end{figure}

In Figure \ref{fig:fluxPA} we show the electron 
intensity versus pitch angle in the energy channels ranging from $0.266$ to 
$39.50$ keV for the shock event on 2000 Feb 11. As shown in the following, 
from observational data we get similar results as YEA2018 did. We also
carry out a large scale numerical simulations for electron acceleration under 
the condition set for the event with the upstream observations as the 
initial distribution.

In the upper pannels of Figures \ref{fig:fluxPA}(a)--(l), the blue and red 
diamonds show the 10-minute averaged observational data upstream and downstream 
of the shock, respectively, and the black circles
show the downstream simulation results. 
It is shown that, except for the highest energy channel shown in Figure
\ref{fig:fluxPA}(l), the 
upstream electron intensities display an anisotropic distribution with 
higher values in the parallel and anti-parallel magnetic field directions 
and lower values in the perpendicular direction. The anisotropy decreases as
the energy increases. In Figure \ref{fig:fluxPA}(l) with the highest energy,
$39.50$ keV, 
the observed upstream intensities are high and low in the parallel and
anti-parallel directions, respectively, and the dramatic change of values happens
around $90^\circ$ pitch angle. 
The downstream simulation results indicate that electron 
intensities increase around $\sim$90$^\circ$ compared with upstream intensities.

In the lower pannels of Figures \ref{fig:fluxPA}(a)--(l), red 
diamonds show the ratio of the downstream to upstream intensities 
for observations. It is shown that from observations for energy channels
except for the largest energy one, the ratio in
perpendicular directions is the largest.
In the lower pannels of Figures 
\ref{fig:fluxPA}(a)--(l) black circles show the ratio of downstream simulation
intensities to upstream observation intensities.
The lower panels of Figure \ref{fig:fluxPA} show that at all energy 
channels the ratio of the downstream to upstream electron intensities 
for both observations and simulations peaks at $\sim$80$^\circ$--100$^\circ$,
with a much clear trend in the lower energy channels of 0.266--4.161 keV in
Figure \ref{fig:fluxPA}(a)--(g). This indicates that at quasi-perpendicular 
shocks there is the strongest acceleration in the pitch angle around $90^\circ$. In addition,
this acceleration is more efficient for particles with lower energies.

\begin{figure}
\epsscale{1.}
\plotone{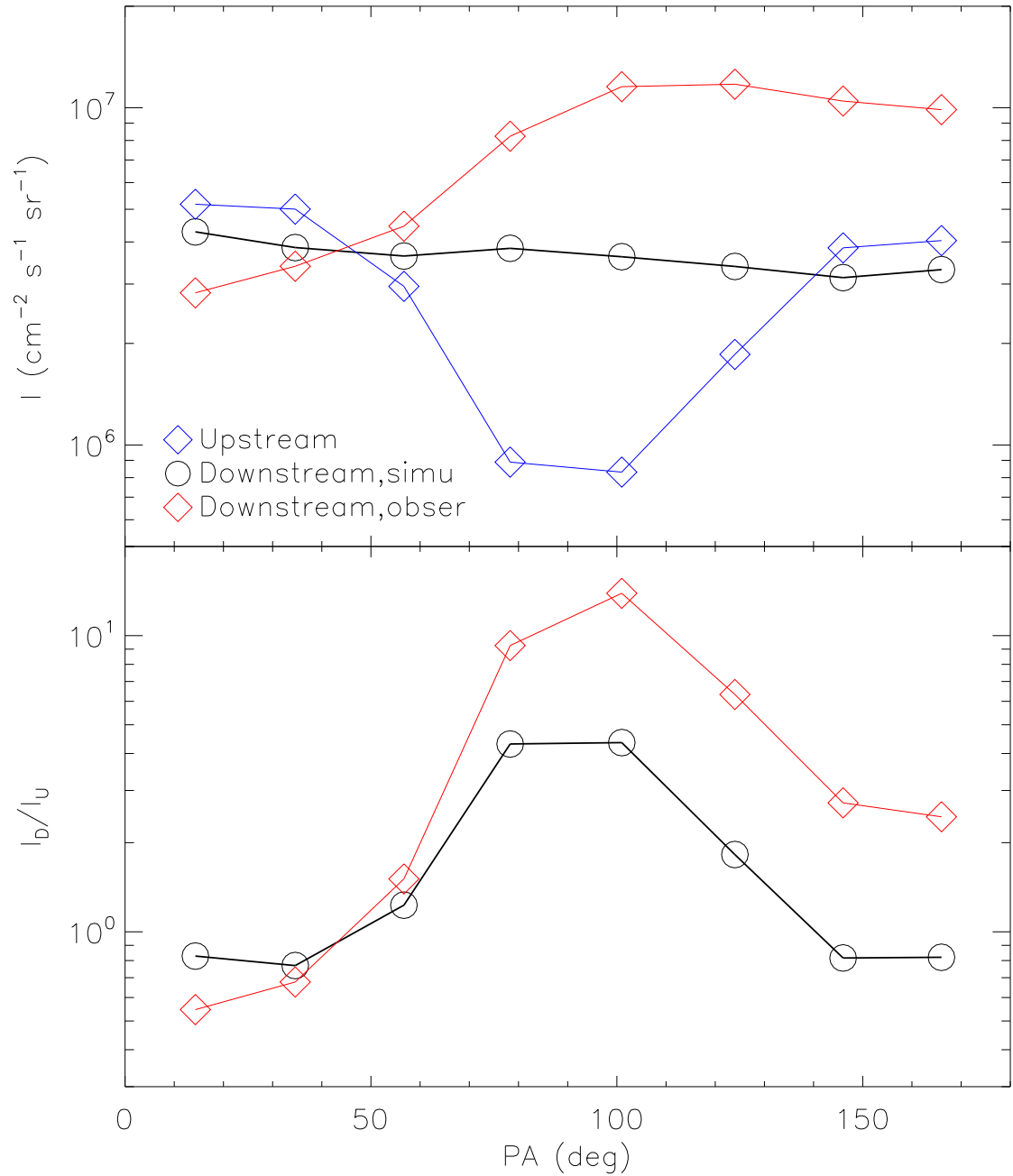}
        \figcaption{Upper panel: Integral over the energy range 0.266--39.50 keV 
of the differential intensity spectrum in each pitch angle direction for the
upstream (blue diamonds), downstream obtained from simulations (black circles),
and downstream from observations (red diamonds). Lower panel: Red diamonds 
(black circles) show the ratio of the downstream integral intensity for 
observations (simulations) to the upstream intensity.
\label{fig:Integral}}
\end{figure}

For each pitch angle direction the integral energy intensities $I_U$ and $I_D$ in the upstream and downstream, respectively, are obtained
by integrating the differential intensity 
over the energy range of 0.266--39.50 keV. In the upper panel of Figure \ref{fig:Integral},
blue and red diamonds show the upstream and downstream observational integral energy
intensities, respectively, and black circles indicate the downstream one from simulations.
From the figure it is shown that, with observations,
the upstream integral energy intensity has the lowest value in around $90^\circ$
pitch angle. However, the downstream integral energy intensity is the highest in
around $90^\circ$ pitch angle, and relatively higher in the anti-parallel direction than
in the parallel direction. In addition, the downstream integral energy 
intensity obtained from simulations, compared with the upstream integral energy intensity,
is higher in around $90^\circ$
pitch angle and lower in the parallel and anti-parallel directions. 
The lower panel of Figure \ref{fig:Integral} shows the ratio of the downstream to
upstream integral energy intensities. It is shown that from both the observations 
(red diamonds) and simulations (black circles)
the ratio reaches its peak around the perpendicular direction. This indicates that the
shock acceleration efficiency is the strongest in the perpendicular direction according
to the observations and simulations.
On the other hand, we can find that the ratio from observations is larger than that
from simulations in the perpendicular direction, which indicates that our simulations do
not produce shock acceleration efficiency as strong as that from observations. Furthermore, 
in the anti-parallel
direction, the ratio from observations is larger than that from simulations. The reason
might be that there are anisotropic beams in the anti-sunward-traveling (anti-parallel)
direction in the downstream of the shock (YEA2018). 

\begin{figure}
\epsscale{1.}
\plotone{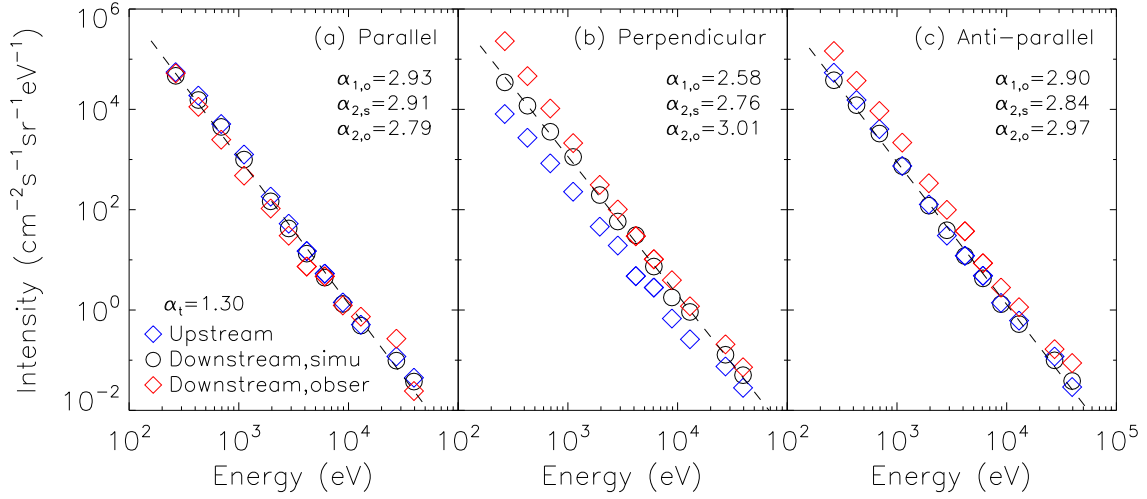}
        \figcaption{Electron energy spectra in the directions parallel (a), 
perpendicular (b), and anti-parallel (c) to the magnetic field. The blue 
(red) diamonds denote 10-minute averaged electron intensities in the upstream
(downstream) of the shock. The black circles correspond to the downstream intensities
obtained from simulations. The values of $\alpha_{1,\text o}$, $\alpha_{2,\text o}$,
and $\alpha_{2,\text s}$ correspond to the energy spectral indices of power-law fits 
to observational data in the upstream and downstream regions, and simulations
in the downstream, respectively. Dashed line indicates the power-law fitting of the
simulation results in the downstream. Also denoted is the theoretical spectral index, 
$\alpha_{\text{t}}$.
\label{fig:spectra}}
\end{figure}

In Figure \ref{fig:spectra} we compare the energy spectra of electrons in the
directions parallel (a), perpendicular (b), and anti-parallel (c) to the magnetic
field. The blue and red diamonds denote 10-minute averaged electron intensities in
the upstream and downstream of the shock, respectively.  The black circles indicate
the simulation results in the downstream. From the figure we can see that the observed 
downstream intensities are several times the upstream intensities in the perpendicular
and anti-parallel directions, but they are similar in the parallel direction. In 
addition, the simulated downstream intensities increase significantly 
relative to the initial upstream intensities, which is set as the upstream 
observations, in the perpendicular direction, but they stay almost unchanged
in the parallel and anti-parallel directions. It is shown that a more prominent 
intensity enhancement occurs in the perpendicular direction for both observations
and simulations, especially in the 
energy channel of 0.266 keV the perpendicular intensity increases 
$\sim 30$ times for observations and $\sim 5$ times for simulations. 
We fit the electron spectrum in each direction as a power law with spectral indices
$\alpha_{1,\text{o}}$, $\alpha_{2,\text{o}}$, and $\alpha_{2,\text{s}}$, where 
subscripts $1$ and $2$ denote upstream and downstream, and o and s denote
observations and simulations, respectively. The dashed line indicates 
the power-law fitting of the 
simulation results in the downstream. We can see that the values of 
$\alpha_{1,\text{o}}$, $\alpha_{2,\text{o}}$, and $\alpha_{2,\text{s}}$ are larger
than $\alpha_{\text{t}}=1.30$ which is predicted by diffusive shock acceleration. 
It is shown that the direction perpendicular to the magnetic field at the quasi-perpendicular 
shock front plays an important role in the shock acceleration of particles.

\begin{figure}
\epsscale{1.}
\plotone{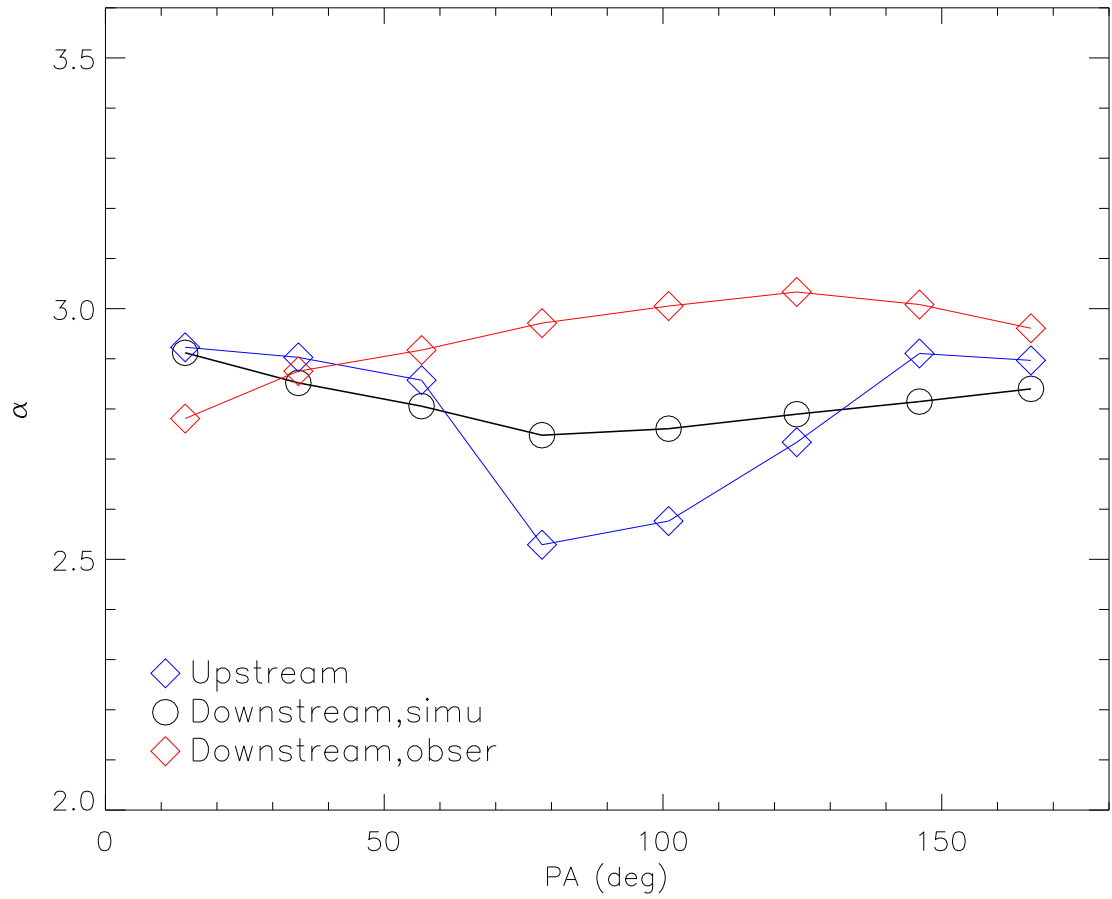}
        \figcaption{Energy spectral index by a power-law fit to the energy spectrum in 
each pitch angle direction in Figure \ref{fig:fluxPA}. The blue and red diamonds 
indicate the results from the observations in the upstream and downstream, respectively. 
The black circles indicate the results from simulations in the downstream.
\label{fig:alphaPA}}
\end{figure}

In addition, we plot the energy spectral index as a function of electron
pitch angle in Figure \ref{fig:alphaPA}. Blue and red diamonds show the results
from the upstream and downstream observations, respectively. We also show the spectral 
indices in the downstream from simulations with black circles. It is shown that the 
downstream spectral indices in the perpendicular direction from both observations
and simulations are larger than the upstream one. It is assumed that
there is more effective shock acceleration for lower energy particles, and
strong shock acceleration causes much higher flux enhancement in the lower energy 
range for energetic particles, generating softer downstream particle spectrum. 
Accordingly, the much softer downstream energy spectrum from both observations
and simulations relative to the upstream observations in the quasi-perpendicular 
direction indicates stronger shock acceleration in this direction. However, the 
observed spectral indices in the downstream are higher than 
those in the upstream in the anti-parallel direction, 
the reason might also be the anti-sunward-travelling beams downstream of the shock
as mentioned above. 

\begin{figure}
\epsscale{1.}
\plotone{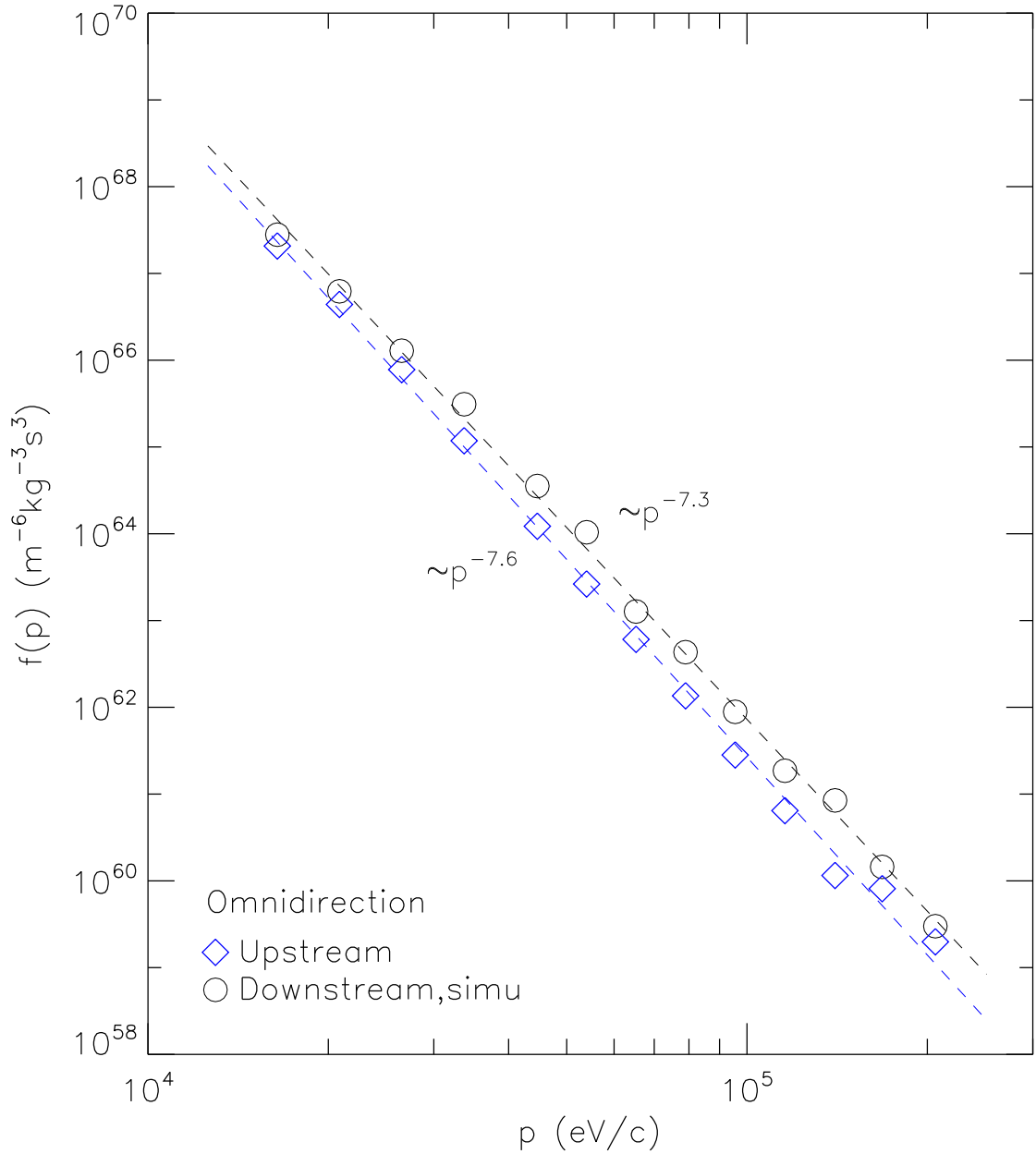}
        \figcaption{Omnidirectional momentum distributions for observations 
in the upstream (blue diamonds) and simulations in the downstream (black circles), 
and their power-law fits with dashed lines. 
\label{fig:distribution}}
\end{figure}

The fact that the acceleration of electrons in the perpendicular direction is 
more efficient reveals the importance of SDA process at quasi-perpendicular
shocks. So that we calculate 
the electron drift length and drift time from the data of distribution functions
using the method in YEA2018. 
As shown in Figure \ref{fig:distribution}, the 
distribution functions, $f_1$ (blue dashed line) and $f_2$ (black dashed line), 
are obtained by a linear fit in log-log space to the data points from 
observations in the upstream (blue diamonds) and simulations in the downstream (black 
circles), respectively. We are able to obtain the energy gain $\Delta E$ 
after the acceleration of upstream electrons with a momentum $p_1$ with Equation (\ref{eq:same_phase_space_density}), considering the same phase space density. 
It is
noted that in this method, the data of distribution functions are used. 
Then, the electron drift length and drift time from simulations 
can be obtained through Equation (\ref{eq:Ldrift}) and Equation (\ref{eq:Tdrift}). 

\begin{figure}
\epsscale{1.}
\plotone{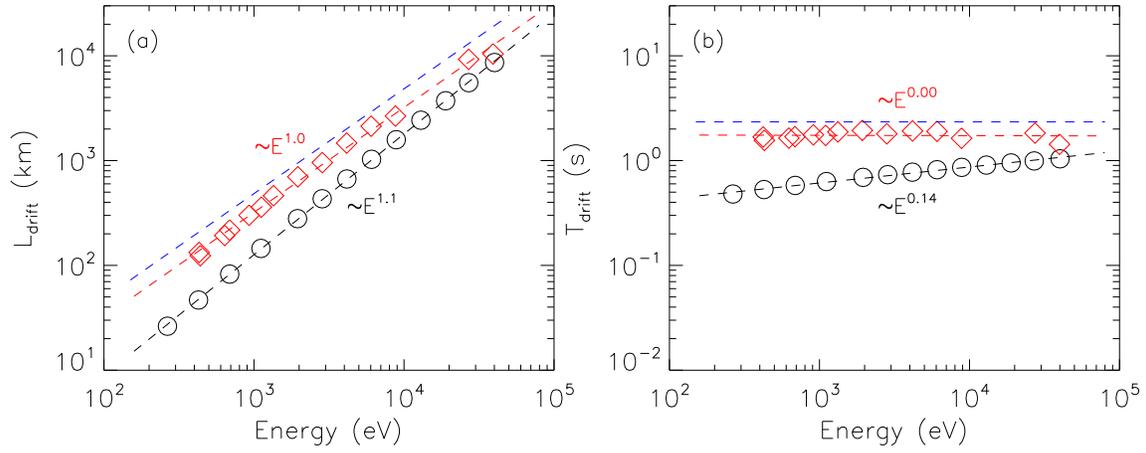}
        \figcaption{
The drift length $L_{\text{drift}}$ and the drift time $T_{\text{drift}}$ as 
a function of electron energy in (a) and (b), respectively. 
Red diamonds and black circles indicate results from observations in
YEA2018 and our simulations, respectively.
Red and black dashed lines indicate the linear fitting of the data from
observations and simulations, respectively.
Blue dashed lines indicate the theoretical results. 
\label{fig:LdriftTdrift}}
\end{figure}

Figures \ref{fig:LdriftTdrift}(a) and (b) show the electron 
drift length $L_{\text{drift}}$ and drift time $T_{\text{drift}}$
as a function of energy, respectively.
In Figure \ref{fig:LdriftTdrift}, red diamonds and black circles indicate
results for observations from YEA2018 and simulations in this work,
respectively. It is shown in Figure \ref{fig:LdriftTdrift}(a) that
for the estimated result obtained from simulations the drift 
length increases linearly with the electron energy in log-log space
with a slope $\sim 1.1$, which compares well with that 
from the observations with a slope $\sim 1.0$.
In Figure \ref{fig:LdriftTdrift}(b), the linear fitting of the 
estimated drift time from simulations and electron energy in log-log space
with a slope $0.14$, also agrees approximately with that from the observations
with a slope $0.00$. In other words, the drift time almost does not vary 
with the energy according to both observations and simulations.
We note that for the same energy channel the estimated drift
length and drift time obtained from simulations in this work are lower than
that from observations in YEA2018, which may due to the less 
efficient acceleration of electrons in our numerical model.

In addition, we use Equations (\ref{eq:Ldrift_th})
and (\ref{eq:Tdrift_th}) to calculate the theoretical results of electron 
drift length $L_{\text{drift}}$ and drift time $T_{\text{drift}}$, respectively.
With the theories, it is found that the electron drift length
is proportional to energy $E$ and the electron drift time is independent
of energy. This is consistent with the observations and simulations  
shown in Figure \ref{fig:LdriftTdrift}(a) and (b). 
In Figure \ref{fig:LdriftTdrift}(a) and (b), blue dashed lines indicate the
theoretical results of $L_{\text{drift}}$ and $T_{\text{drift}}$ from Equations 
(\ref{eq:Ldrift_th}) and (\ref{eq:Tdrift_th}), respectively. We can see 
that the theoretical results agree well with the observational
ones, but they are several times larger than the simulation ones.
The reason of the discrepancy might
be that less efficient acceleration is achieved in our simulations.

\section{discussion and conclusions}  

We have used test-particle numerical simulations of a backward-in-time test-particle method 
to study the acceleration of suprathermal electrons in the energy range of 
$\sim$0.3--40 keV at a quasi-perpendicular shock event 
on 2000 February 11 to compare with the observational study in 
YEA2018. We obtain electron pitch angle distributions from
simulations for all 12
energy channels, and find that the ratio of the downstream to upstream 
differential intensities peaks at about $\sim$$90^\circ$ pitch angle.
These results are in good agreement with the spacecraft observations in YEA2018.
In addition, it is found that the observed and simulated electron
energy spectral index for each pitch angle direction in the downstream is 
significantly larger than the theoretical index of diffusive shock acceleration.
The results indicate that shock drift acceleration plays an important role in
the acceleration of electrons at quasi-perpendicular shocks as 
suggested by YEA2018. 

Furthermore, with Liouville’s theorem, considering SDA, YEA2018 used 
observational data to show that the electron drift length $L_{\text{drift}}$
is approximately proportional to the electron energy $E$. In addition, it is
suggested that the drift time  $T_{\text{drift}}$ is almost independent
of the electron energy. From our simulations we obtain the similar 
energy dependence of the drift length and drift time, but with values several
times lower than those from the observations.
Next, we provide theoretical models based on shock drift acceleration
to describe the electron drift length and drift time, which agree well with 
the observational results. In addition, the theories can be used to
explain the energy dependence of electron drift length and drift time
found by observations and simulations. 

It is suggested that the difference between the results from the observations 
and simulations show that our numerical model does not provide shock acceleration 
as effective as the observations. In addition, there is anti-sunward-travelling
beams of energetic electrons downstream of the shock in the observations which does not 
appear in our shock acceleration model.

\acknowledgments
This work was supported, in part, under grants NNSFC
41874206 and NNSFC 41574172.
The work was carried out at National Supercomputer Center 
in Tianjin, and the calculations were performed on TianHe-1 (A).
We gratefully acknowledge
Dr. Linghua Wang for useful discussions about this topic.

\clearpage

\clearpage

\clearpage

\clearpage

\clearpage

\clearpage

\clearpage
\begin{table}
\begin{center}
\caption {Input Parameters for the Shock\label{shockpara}}
\begin{tabular} {ccc}\hline\hline
Parameter & Description & Value \\\hline
$\theta_{\text{Bn}}$ & shock angle & 89$^\circ$ \\
$V_{\text{sh}}$ & shock speed & 682 km s$^{-1}$\\
$s$     & compression ratio &  2.87 \\
$B_{01}$   & upstream magnetic field &  7.0 nT \\
$L_{\text{th}}$ & shock thickness & 2$\times10^{-6}$ AU \\
$U_1$   & upstream speed &  248 km s$^{-1}$ \\
$M_{\text{A1}}$ &upstream Alv$\acute{\text e}$n Mach number & 3.70\\\hline
\end{tabular}
\end{center}
\end{table}

\clearpage
\begin{table}
\begin{center}
\caption {Input Parameters for the Turbulence\label{turbupara}}
\begin{tabular} {ccc}\hline\hline
Parameter & Description &  Value \\\hline
$\lambda$  & slab correlation length & 0.02 au   \\
$\lambda_x$  & 2D correlation length & $\lambda/2.6$   \\
$E_{\text{slab}}:E_{\text{2D}}$ &two-component energy density ratio &
$20:80$\\
${\left(b/B_0\right)^2}_1$ & upstream turbulence level & 0.25    \\
${\left(b/B_0\right)^2}_2$ & downstream turbulence level & 0.36    \\
$k_{\text b}$ & break wavenumber & 10$^{-6}$ m \\
$\beta_{\text i}$ & inertial spectral index & 5/3\\
$\beta_{\text d}$ & dissipation spectral index & 2.7\\\hline
\end{tabular}
\end{center}
\end{table}

\end{document}